\documentclass[trackchanges,twocolumn]{aastex7}
\usepackage{booktabs}
\usepackage{amsmath}
\usepackage{threeparttable} 
\usepackage{tabularx}

\begin{document}

\title{Contribution of Globular Clusters to Diffuse Gamma-ray Emission from Galactic Plane}

\author{Jiayin He}
\affiliation{Key Laboratory of Dark Matter and Space Astronomy, Purple Mountain Observatory, Chinese Academy of Sciences, Nanjing 210023, China}
\affiliation{School of Astronomy and Space Science, University of Science and Technology of China, Hefei 230026, China}
\email{hejy@pmo.ac.cn}

\author{Houdun Zeng}
\affiliation{Key Laboratory of Dark Matter and Space Astronomy, Purple Mountain Observatory, Chinese Academy of Sciences, Nanjing 210023, China}
\email{zhd@pmo.ac.cn}

\author{Yi Zhang}
\affiliation{Key Laboratory of Dark Matter and Space Astronomy, Purple Mountain Observatory, Chinese Academy of Sciences, Nanjing 210023, China}
\affiliation{School of Astronomy and Space Science, University of Science and Technology of China, Hefei 230026, China}
\email{zhangyi@pmo.ac.cn}

\author{Xiaoyuan Huang}
\affiliation{Key Laboratory of Dark Matter and Space Astronomy, Purple Mountain Observatory, Chinese Academy of Sciences, Nanjing 210023, China}
\affiliation{School of Astronomy and Space Science, University of Science and Technology of China, Hefei 230026, China}
\email{xyhuang@pmo.ac.cn}

\author{Qiang Yuan}
\affiliation{Key Laboratory of Dark Matter and Space Astronomy, Purple Mountain Observatory, Chinese Academy of Sciences, Nanjing 210023, China}
\affiliation{School of Astronomy and Space Science, University of Science and Technology of China, Hefei 230026, China}
\email{yuanq@pmo.ac.cn}

\begin{abstract}

The diffuse Galactic $\gamma$-ray emission (DGE) provides a valuable probe for investigating the cosmic ray propagation and interactions within our Galactic environment. Recent observations have demonstrated systematic excesses of DGE compared with the conventional cosmic-ray propagation model predictions. While $\gamma$-ray emissions have been detected in a subset of globular clusters, their undetected populations may significantly contribute to the DGE. Motivated by this possibility, we present a comprehensive assessment of potential contributions from unresolved globular clusters to the DGE. In our analysis, a nonparametric method is employed to estimate the luminosity function and spatial distribution function of globular clusters using the Fermi-LAT fourth source catalog (4FGL) combined with a reference globular cluster catalog. Based on these distributions, we calculate the cumulative contribution of unresolved globular cluster populations to the DGE observed by Fermi-LAT and the Large High Altitude Air Shower Observatory (LHAASO). Our results reveal that globular clusters account for only $\sim$2\% of the DGE at the TeV range, and smaller than $1\%$ in the GeV regime, which is effectively negligible.

\end{abstract}

\keywords{\uat{Globular clusters}{656} --- \uat{Gamma-ray astronomy}{628} --- \uat{Diffuse radiation}{383} --- \uat{Cosmic background radiation}{317}}

\section{Introduction} \label{sec:introduction}

The diffuse Galactic $\gamma$-ray emission (DGE) arises from interactions between cosmic rays and gas and radiation fields in the interstellar medium, involving processes such as neutral pion decay, inverse Compton scattering, and bremsstrahlung \citep{Aharonian_1996, Strong_2000, Strong_2004}. DGE measurements provide crucial insights into cosmic ray propagation within the Milky Way and the nature of the interstellar medium, complementing direct cosmic ray measurements near the solar vicinity.

The DGE was first detected by the OSO-3 satellite \citep{OSO-3_Diffuse_1968, OSO-3_Diffuse_1972}, with subsequent observations from space missions such as SAS-2 \citep{SAS-2_Diffuse_1975}, COS-B \citep{COS-B_Diffuse_1982}, COMPTEL \citep{COMPTEL_Diffuse_1994, COMPTEL_Diffuse_1996}, HEAO-1 \citep{HEAO-1_Diffuse_1997}, and EGRET \citep{EGRET_Diffuse_1997}. The space detector Fermi-LAT extended the measurements to $O(100)$ GeV, and a comprehensive comparison between the measurements and the propagation model predictions was conducted  \citep{FermiLAT_Diffuse_2012}. Observations of DGE from TeV to PeV range have been conducted via ground-based experiments like Milagro \citep{Milagro_Diffuse_2007, Milagro_Diffuse_2008}, HESS \citep{HESS_Diffuse_2014}, ARGO-YBJ \citep{ARGO_Diffuse_2015}, Tibet AS\(\gamma\) \citep{ASgamma_Diffuse_2021}, LHAASO \citep{KM2A_Diffuse_2023,WCDA_Diffuse_2025}, and HAWC \citep{HAWC_Diffuse_2024}.

Recent measurements from LHAASO and HAWC across a broad energy range have indicated that conventional cosmic ray propagation models may underestimate DGE. LHAASO observations reveal that fluxes from the inner Galactic region ($15^{\circ}<l<125^{\circ}$, $-5^{\circ}<b<5^{\circ}$) exceed model predictions by a factor of 3, whereas in the outer Galactic region ($125^{\circ}<l<235^{\circ}$, $-5^{\circ}<b<5^{\circ}$), the discrepancy is by a factor of 2 \citep{KM2A_Diffuse_2023, WCDA_Diffuse_2025}. HAWC data from the region $43^{\circ}<l<73^{\circ}$, $-5^{\circ}<b<5^{\circ}$ shows DGE fluxes that are higher by a factor of 2 compared with the DRAGON-based models \citep{HAWC_Diffuse_2024}. The model calculated DGE was also found to be lower than the data above several GeV in the region of $-80^{\circ}<l<80^{\circ}$, $-8^{\circ}<b<8^{\circ}$ when compared with the measurement of Fermi-LAT \citep{FermiLAT_Diffuse_2012}. Recent re-analysis of diffuse emission from Fermi-LAT in the region of interest (ROI) of LHAASO also suggests the need for an additional component to bridge the gap between observed fluxes and model calculations \citep{Zhang_2023, WCDA_Diffuse_2025}.

Several models have been proposed to explain the DGE excess, such as various kinds of unresolved sources including pulsar halos or pulsar wind nebulae \citep{universe9090381,Yan_2024,PhysRevD.109.083026,Chen_2024}, X-ray binaries \citep{yue2024xraybinariespotentialdominant,kuze2025multimessengeremissionmagneticallyarrested}, massive star clusters \citep{Menchiari_2025}, confinement and interaction around cosmic accelerators \citep{PhysRevD.110.103039,He_2024,Nie_2024,yang2024confinementrelativisticparticlesvicinity,ambrosone2025originveryhighenergydiffusegammaray}, or cosmic ray interactions in alternative propagation setup \citep{Giacinti:2023ljr,luque2025cosmicrayseaexplainsdiffuse,marinos2025temporalvariabilitygalacticmultitev}. In this work, we discuss the contribution of unresolved globular clusters to DGE, which likely contain a large population of millisecond pulsars (MSPs). 

Over thirty globular clusters have been detected in the GeV band in the 4FGL catalog \citep{4FGL_2020, 4FGL_DR4_2024}, where MSPs are the primary emission sources, as supported by successful MSP detections in globular clusters \citep{MSPs_GLC_FermiLAT_2011, MSPs_GLC_FermiLAT_2013}. At higher energies, HESS has detected TeV $\gamma$-ray emissions from Terzan 5 \citep{Terzan5_HESS_2011}. The emission mechanisms in globular clusters may include curvature radiation (CurvR) from MSP magnetospheres and inverse Compton (IC) scattering from e\(^\pm\) pairs. \citet{GLCs_Song_2021} identified evidence for a high-energy tail in $\gamma$-ray spectra of globular clusters, which could be due to the IC emission. \citet{Terzan5_ICModel_2024} also modeled Terzan 5's very-high-energy $\gamma$-ray emission detected by HESS as the IC process. It is thus possible that globular clusters are emitters of $\gamma$-ray emission above TeV energies in general. According to the Harris catalog (2010 edition), there are 157 Milky Way globular clusters \citep{Harris_GLCs_1996}, yet fewer than a quarter are detected in the GeV range and only one in the TeV range, suggesting that unresolved globular clusters could meaningfully contribute to the DGE.

This study focuses on the potential contributions of unresolved globular clusters to the DGE. The paper is structured as follows: Section \ref{sec:data} provides a brief description of our data sample. Section \ref{sec:distribution} discusses the intrinsic distribution of sources using a non-parametric statistical method. Section \ref{sec:est_unresol} gives the calculation method and results on the DGE contribution from unresolved globular clusters. Finally, Section \ref{sec:conclusion_discussion} presents our conclusion and a discussion.

\section{Data sample} \label{sec:data}

In this analysis, we utilize globular clusters that are detected with a significance greater than 5\(\sigma\) in the Fermi-LAT 14-year source catalog, data release 4 (4FGL-DR4), totaling 36 sources \citep{4FGL_DR4_2024}. These sources are measured over an energy range from 50 MeV to 1 TeV and follow either a log-parabola or power-law spectral form. Notably, only three sources favor a power-law spectrum over a log-parabola. For simplicity, all globular clusters in this study are modeled with a log-parabola spectrum. Using distance parameters from the Harris catalog (2010 edition) \citep{Harris_GLCs_1996}, we compute the luminosity of these globular clusters as
\begin{equation}\label{eq:luminosity}
	L = 4\pi D^{2} \int _{E_\mathrm{min}} ^{E_\mathrm{max}} \phi_{0} \left( \frac{E}{E_0} \right) ^{-\Gamma_{1} -\Gamma_{2} \, \mathrm{ln} (E/E_0) } E\,dE,
\end{equation}
where \(D\) is the distance from the globular cluster to the Earth. The spectral parameters \((\phi_{0}, E_0, \Gamma_{1}, \Gamma_{2})\) are obtained from the 4FGL-DR4 fits, with integration bounds \(E_\mathrm{min} = 50~\mathrm{MeV}\) and \(E_\mathrm{max} = 1~\mathrm{TeV}\).

\begin{figure}[!htbp]
\centering
\includegraphics[width=0.45\textwidth]{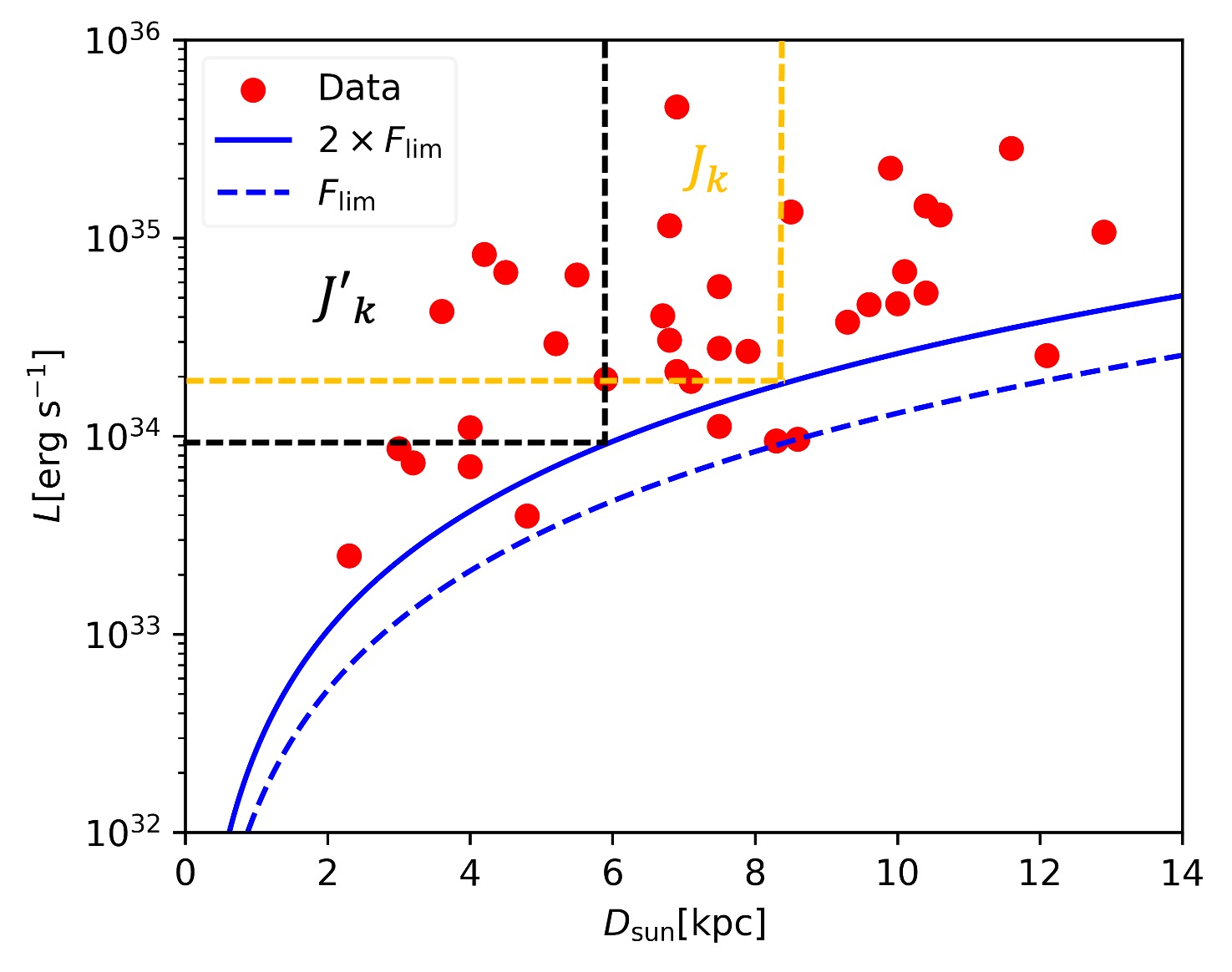}
\caption{The luminosities and distances of Fermi-LAT detected globular clusters (red dots). The luminosity is estimated using Equation \textbf{(\ref{eq:luminosity})} across an energy range from 50 MeV to 1 TeV. The blue dashed line indicates luminosity calculated using the minimum integrated energy flux \(F_\mathrm{lim}\), with the blue solid line representing twice this minimum luminosity. The orange dashed square and black dashed square define the data sets \(J_{k}\) and \(J'_{k}\) (see Section 3), respectively.}
\label{fig:data}
\end{figure}

Due to the limited sensitivity of the instrument, only sources with a flux above a specific threshold can be detected. This threshold is determined by the minimum energy flux, \(F_\mathrm{lim}\), given by
\begin{equation}
F_\mathrm{lim}= \left[\,\int _{E_\mathrm{min}} ^{E_\mathrm{max}} \phi_{0,i} \left( \frac{E}{E_{0,i}} \right) ^{-\Gamma_{1,i} -\Gamma_{2,i} \, \mathrm{ln} (E/E_{0,i}) } E\,dE \,\right] _\mathrm{min}.
\end{equation}
It corresponds to a minimum luminosity of \(L_\mathrm{lim}=4\pi D^{2}F_\mathrm{lim}\). The estimated minimum luminosity is shown as a blue dashed line in Figure \ref{fig:data}. The blue solid curve in Figure \ref{fig:data} represents twice the minimum luminosity. As can be seen in this Figure, the luminosity of the observed globular clusters ranges from approximately \(10^{33}\) to \(10^{36}\, \mathrm{erg~s^{-1}}\), with distances less than 14 kpc.

\begin{figure*}[ht!]
\centering
\includegraphics[width=1.0\textwidth]{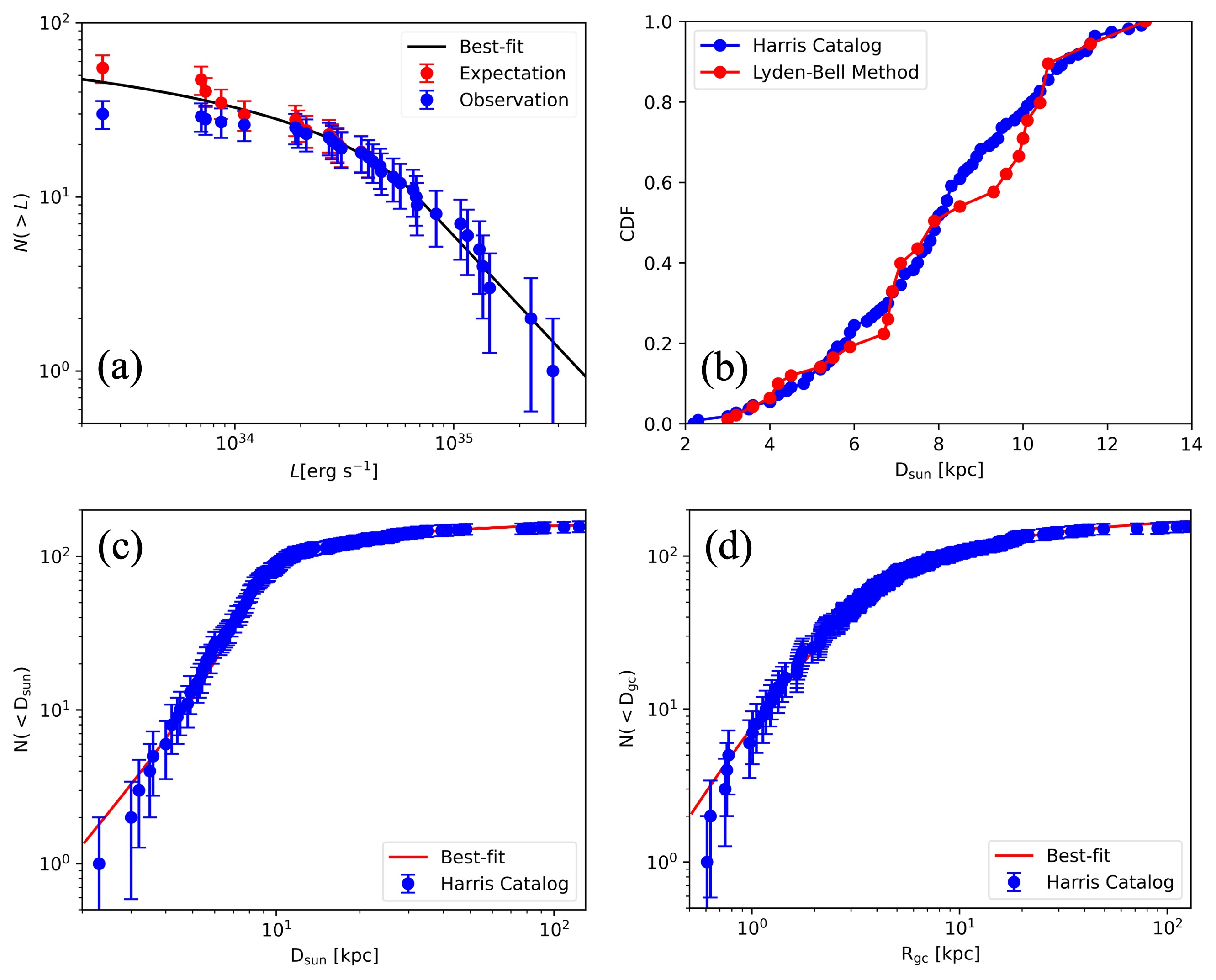}
\caption{Intrinsic cumulative distributions of luminosity and spatial density. (a): The intrinsic cumulative distribution of luminosity (red data points) derived by the Lyden-Bell C$^-$ method \citep{LedynBell_1971} along with counts from Fermi-LAT detected sources (the blue data points) in Figure \ref{fig:data}, while the black curve shows the best fit to the red points; (b): K-S test for the distribution derived by Lyden-Bell method and that from Harris catalog \citep{Harris_GLCs_1996}; Fitting to the distance distribution of $D_\text{sun}$ (c) and the distance distribution of $R_\text{gc}$ (d) with Equation (\ref{eq:fDlb}).}
\label{fig:cdfs}
\end{figure*}

\section{Intrinsic Distribution of Sources} \label{sec:distribution}
In this section, we employ the Lynden-Bell C\(^-\) method \citep{LedynBell_1971} to derive the intrinsic distribution of sources. This non-parametric approach has been utilized to estimate the intrinsic distribution of integrated flux and photon index for ultra-high energy $\gamma$-ray sources from a truncated sample using the LHAASO First Catalog \citep{UnresolvedSources_hejy_2025}. In our analysis, we assume that the luminosity and distance are independent variables, and any correlation introduced by selection effects is removed by applying the Lynden-Bell C\(^-\) method.

\subsection{Luminosity distribution} \label{sucsec:cdf_l}

To compute the intrinsic luminosity distribution of globular clusters, we first define an associated set for the \(k\)-th source as
\begin{equation}
	J_{k} = \{ j | L_{j} \geq L_{k}, D_{\mathrm{sun},j} \leq D ^\mathrm{max} _{\mathrm{sun},k} \}.
\end{equation}
This dataset, represented by the orange dashed square in Figure \ref{fig:data}, includes sources with luminosities no less than that of the \(k\)-th source and with distances smaller than the farthest source at or below the \(k\)-th source's luminosity. For the analysis, we select the twice-the-minimum luminosity curve as a conservative detection threshold, shown as the blue solid curve in Figure \ref{fig:data}. This approach is similar to that taken in \citep{petrosian2015cosmological}, and we have verified that the results are consistent with those derived using the minimum luminosity curve. The number of sources within the set \(J_{k}\) is denoted as \(n_{k}\).

Using the Lynden-Bell C\(^-\) method \citep{LedynBell_1971}, the cumulative luminosity distribution can be derived as
\begin{equation}\label{eq:lbc_cdfl}
	\psi(L_{k})=\prod _{j} \left( 1+ \frac{1}{n_{j}-1} \right),
\end{equation}
where the operation is over sources with luminosities \(L_{j} > L_{k}\), and \(n_{j}-1\) represents the number of sources in the set \(J_{j}\), excluding the \(k\)-th source. From Equation \textbf{(\ref{eq:lbc_cdfl})}, \(n_{j}-1\) represents the total number of sources with \(L_{j}>L_{k}\). The cumulative distribution can be expressed in a differential form of \(dN/dL\):
\begin{equation}\label{eq:dndl}
	\psi(L) \equiv N(>L)=\int ^{\infty} _{L} \frac{dN}{dL'}dL'\,.
\end{equation}
\vspace{0.5em}

In our analysis, a broken power law is utilized to describe the differential distribution of the luminosity function:
\begin{equation}\label{eq:l_diff}
	\frac{dN}{dL} = \left\{ \begin{array}{ll}
		A \left( \frac{L}{L_{0}} \right)^{-\beta_{1}} & \text{for } L < L_{b}, \\
		A \left( \frac{L_{b}}{L_{0}} \right)^{-\beta_{1} + \beta_{2}} \left( \frac{L}{L_{0}} \right)^{-\beta_{2}} & \text{for } L \geq L_{b},
	\end{array}\right.
\end{equation}
where \(A\) is the normalization at the pivot luminosity \(L_0 = 10^{34}\,\mathrm{erg\,s^{-1}}\); \(\beta_{1}\) and \(\beta_{2}\) are the spectral indices below and above the break luminosity \(L_{b}\), respectively. In Equation \textbf{(\ref{eq:l_diff})}, \(A\) is not crucial because \(dN/dL\) will be normalized to unity when estimating the flux from globular clusters in Section \ref{sec:est_unresol}.

In plot (a) of Figure \ref{fig:cdfs}, we present the intrinsic cumulative distribution derived from the data using the Lynden-Bell C\(^-\) method (red points), and compare it with the distribution obtained by directly counting observed sources above the blue solid curve. The black solid curve represents the best-fit model from the data, with parameters and errors summarized in Table \ref{tab:fittings}.

\begin{table*}[htb!]
\begin{threeparttable} 
\centering
\caption{Fitting results of the intrinsic distributions of luminosity and spatial density.}
\label{tab:fittings}
\begin{tabular}{ccccccccccc}
\toprule
 Parameter & $\beta_{1}$ & $\beta_{2}$ & $\log_{10} L_{b}$ & $N_{0}$ & $a$ & $b$ &  $r_b$ & $z_\text{h}$ &$\Gamma_{1}$ & $\Gamma_{2}$  \\             
\midrule
Mean & 0.859  &  2.342 & 34.833 & 0.977 &  0.521  & 2.458 & 2.951 & 0.379 & 2.122 & 0.270 \\
Error &0.690  &   0.416 &  0.308 & 0.510 & 0.135 & 0.081 &  0.272 &  0.117 & 0.196 &  0.055 \\
\bottomrule
\end{tabular}
\vspace{1em}
\begin{tablenotes}
\footnotesize
\item\textbf{Note.} In our analysis, $L_{0}=10^{34}\,\mathrm{erg\,s^{-1}}$, and $E_{0}=1\,\mathrm{GeV}$. $L_{b}$ is in units of $\mathrm{erg\,s^{-1}}$, $r_b$ and $z_\text{h}$ are in units of kpc, and $N_{0}$ is in units of $\mathrm{kpc^{-3}}$.
\end{tablenotes}
\end{threeparttable} 
\end{table*}

\subsection{Spatial density distribution} \label{sucsec:cdf_d}

The calculation for the intrinsic spatial distribution of globular clusters is similar to those described above. We define a data set \(J'_{k}\) for the \(k\)-th source as
\begin{equation}
    J'_{k} = \{ j | L_{j} > L ^\mathrm{lim} _{k}, D_{\mathrm{sun},j} \leq D  _{\mathrm{sun},k} \},
\end{equation}
where \(L ^\mathrm{lim}_{k}\) denotes the minimum luminosity given the distance of the \(k\)-th source. This dataset is depicted by the black dashed square in Figure \ref{fig:data}, and the number of sources within this set is \(m_{k}\). The number of sources with distances less than \(D_{\mathrm{sun},k}\) is calculated by
\begin{equation}\label{eq:lbc_cdfd}
    \psi(D_{\mathrm{sun},k})= \prod _{j} \left( 1+ \frac{1}{m_{j}-1} \right),
\end{equation}
where the product is taken over all sources with \(D_{\mathrm{sun},j}<D_{\mathrm{sun},k}\). We compare this cumulative distribution with the one derived from direct counting of 157 globular clusters listed in the Harris catalog \citep{Harris_GLCs_1996}, as shown in plot (b) of Figure \ref{fig:cdfs}. The cumulative distribution from the Harris catalog is normalized to the maximum distance of globular clusters detected by the Fermi-LAT. To assess the consistency between the two distributions, we perform the Kolmogorov-Smirnov (K-S) test, which indicates a maximum difference of $|\text{CDF}_\text{Lyden-Bell} - \text{CDF}_\text{Harris}| = 0.12$ at a distance of 9.6 kpc from the Sun, based on a sample size of 31 (represented by the red points above the blue solid curve in Figure \ref{fig:data}). The resulting p-value is 0.72, suggesting that the distribution obtained using the Lyden-Bell method is in good agreement with that estimated from the Harris catalog. In subsequent analyses, we use the Harris sample to compute the spatial distribution function.

The number of sources within distance \(D_\text{s}\) can be expressed by
\begin{equation}\label{eq:fDlb}
    N(<D_\text{s})=\int ^{2\pi} _{0} dl \int ^{\frac{\pi}{2}} _{-\frac{\pi}{2}}  \mathrm{cos}b\,db  \int ^{D_\text{s}} _{0} f(D,l,b)D^{2}dD,
\end{equation}
where \( \mathbf{f(D,l,b)} \) is the spatial density function of globular clusters expressed in the Galactic coordinate, $D$ is the distance to the globular cluster from the solar system. We assume that the form can be expressed as separable functions of radial distance \( r \) and height \( z \), such that \( f(r,z) = f(r)g(z) \). In our analysis, \( f(r) \) is modeled as a broken power law, while \( g(z) \) follows an exponential decline. This assumption accounts for the density variation with Galactocentric distance \( R_\text{gc} \), characterized by different indices in the regions smaller and greater than $\sim$4 kpc \citep{1985ApJ...293..424Z,Harris2001}. Additionally, the number of globular clusters decreases rapidly in the \( z \)-direction \citep{1989AJ.....97..375A,Harris2001}. The complete expression for the spatial density is given by
\begin{equation}\label{eq:dndv}
    f(r,z)=N_{0}\mathrm{e}^{-\frac{|z|}{z_\text{h}}} \left\{ \begin{array}{ll}
		\left( \frac{r}{r_{0}} \right)^{-a} & \text{for } r < r_{b}, \\
		\left( \frac{r_{b}}{r_{0}} \right)^{-a + b} \left( \frac{r}{r_{0}} \right)^{-b} & \text{for } r \geq r_{b},
	\end{array}\right.
\end{equation}
where \( a \) and \( b \) are the indices before and after the breakpoint at \( r_b \), respectively. The parameter \( z_\text{h} \) represents the scale height, \( N_0 \) is the normalization constant, \( r_0 = 8 \, \mathrm{kpc} \), and the density is measured in units of \(\mathrm{kpc}^{-3}\).

To estimate the parameters, Eq. \ref{eq:fDlb} is fitted to the distributions of \( D_\text{sun} \) and \( R_\text{gc} \) from the Harris catalog, as shown in plots (c) and (d) of Figure \ref{fig:cdfs}. The fitting results are summarized in Table \ref{tab:fittings}.

\begin{figure}[!htbp]
\centering
\includegraphics[width=0.45\textwidth]{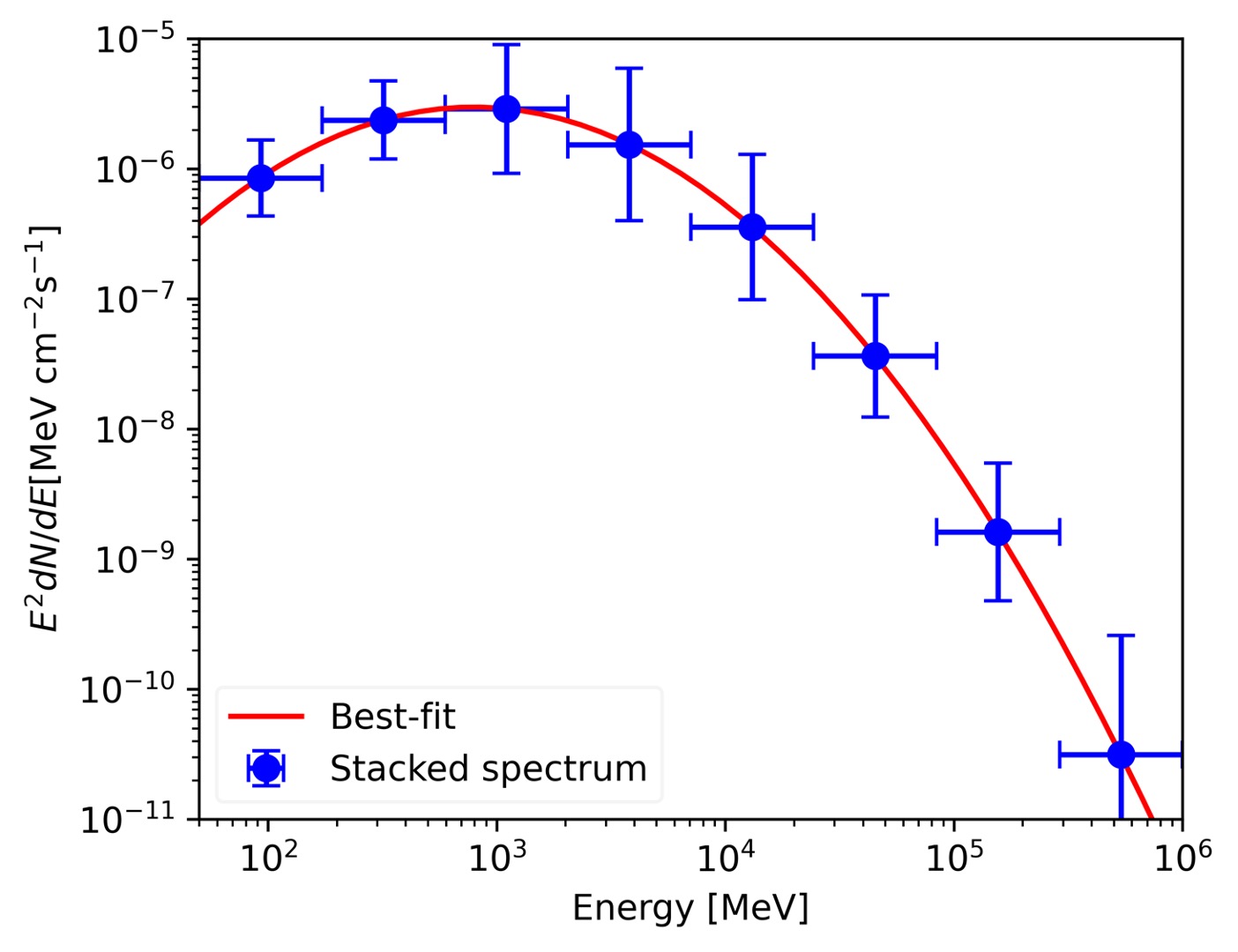}
\caption{ The Stacked SED of globular clusters with integrated flux above $1.6\times 10^{-8}\,\mathrm{ph\,cm^{-2}s^{-1}}$ along with the best-fit. The error bars represent the standard deviation of the flux distribution in each bin.}
\label{fig:spec}
\end{figure}

\section{Estimation of fluxes from unresolved globular clusters} \label{sec:est_unresol}
We have derived the intrinsic luminosity and spatial distributions from the data. To estimate the flux of the globular cluster population, it is crucial to determine their spectra. In this section, we first estimate the spectra of the globular clusters. Using these spectra along with the previously obtained distributions, we then compute the flux from the ROI, which matches the experimental settings.

\begin{figure*}[!htbp]
\centering
\includegraphics[width=1.0\textwidth]{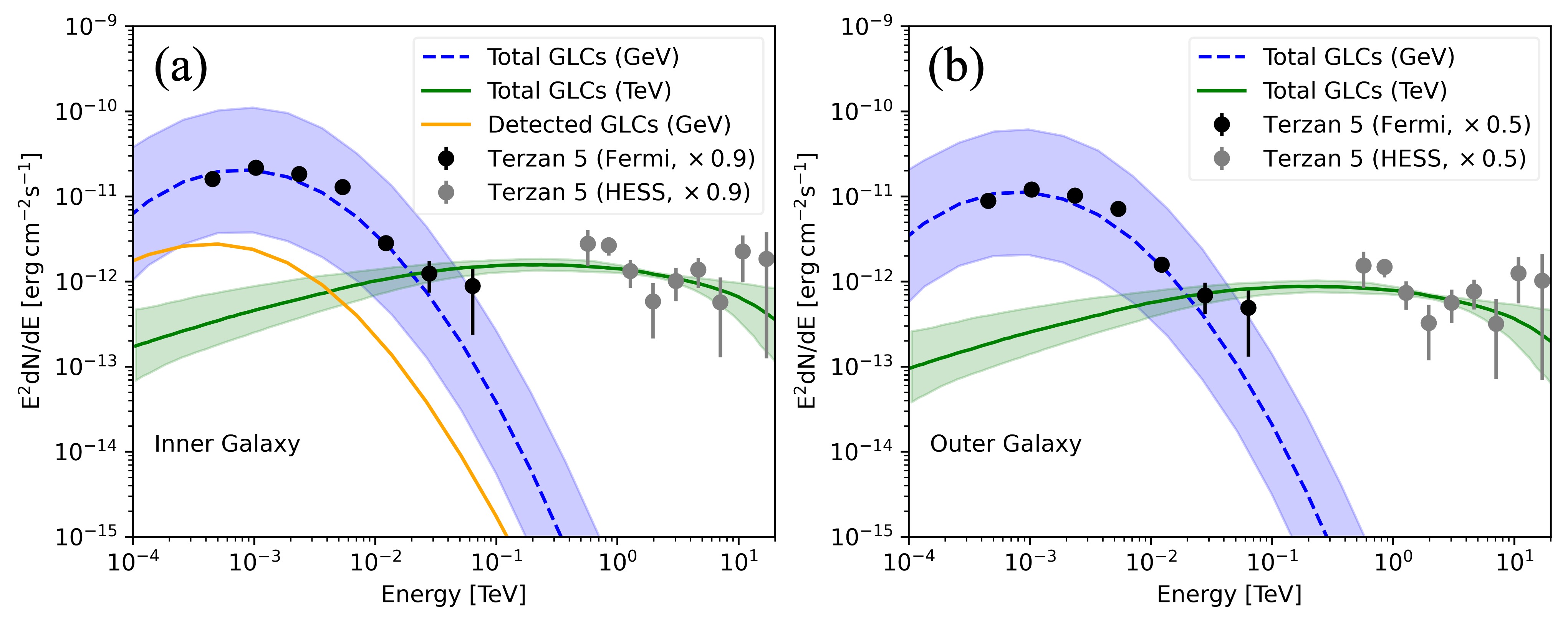}
\caption{The SEDs of total contribution from globular clusters in the inner Galaxy region (panel (a)) and outer Galaxy region (panel (b)). The black points are measured fluxes from Terzan 5 \citep{Terzan5_ICModel_2024,GLCs_Song_2021}. In both plots, the green bands represent TeV component emissions, derived from the SED of Terzan 5's modeled IC emission \citep{Terzan5_ICModel_2024} and scaled by factors of 0.9 and 0.5, respectively.}
\label{fig:scale_terzan5}
\end{figure*}

\subsection{Intrinsic spectrum}

To derive the intrinsic spectra of globular clusters, we perform a stacking analysis, selecting sources with sufficiently large integrated fluxes to create a flux-limited sample, thereby removing selection effects \citep{FermiLAT_EDB_2010}. The log-parabola spectra are integrated over an energy range from 50 MeV to 1 TeV to obtain the integrated flux, defined as $F=\int _{50\,\mathrm{MeV}}  ^{1\,\mathrm{TeV}} \phi_{0} \left( \frac{E}{E_0}
\right) ^{-\Gamma_1 - \Gamma_{2}\,\mathrm{ln}(E/E_0)}dE$. We find that the stacked spectrum from sources with an integrated flux of approximately \(1.6 \times 10^{-8}\,\mathrm{ph\,cm^{-2}s^{-1}}\) remains stable when making minor adjustments to the flux threshold.

To estimate the stacked spectrum, we calculate the flux for each globular cluster with an integrated flux above \(1.6 \times 10^{-8}\,\mathrm{ph\,cm^{-2}s^{-1}}\) using spectrum parameters from 4FGL-DR4. The average flux for each energy band is calculated as the arithmetic mean of all globular cluster fluxes within that band. The resulting stacked spectrum, as shown in Figure \ref{fig:spec}, is fitted with a log-parabola using a pivot energy of 1 GeV. The error bars in Figure \ref{fig:spec} indicate the standard deviation of the flux distribution, serving as a conservative estimate of the uncertainty in the average flux. The best-fit spectrum is represented by the red curve, and the spectral fitting information is summarized in Table \ref{tab:fittings}. While the emission could be related to CurvR, we refer to it as the GeV component emission, as this study does not delve into the specific origin of the emission.

\begin{figure*}[!htbp]
\centering
\includegraphics[width=1.0\textwidth]{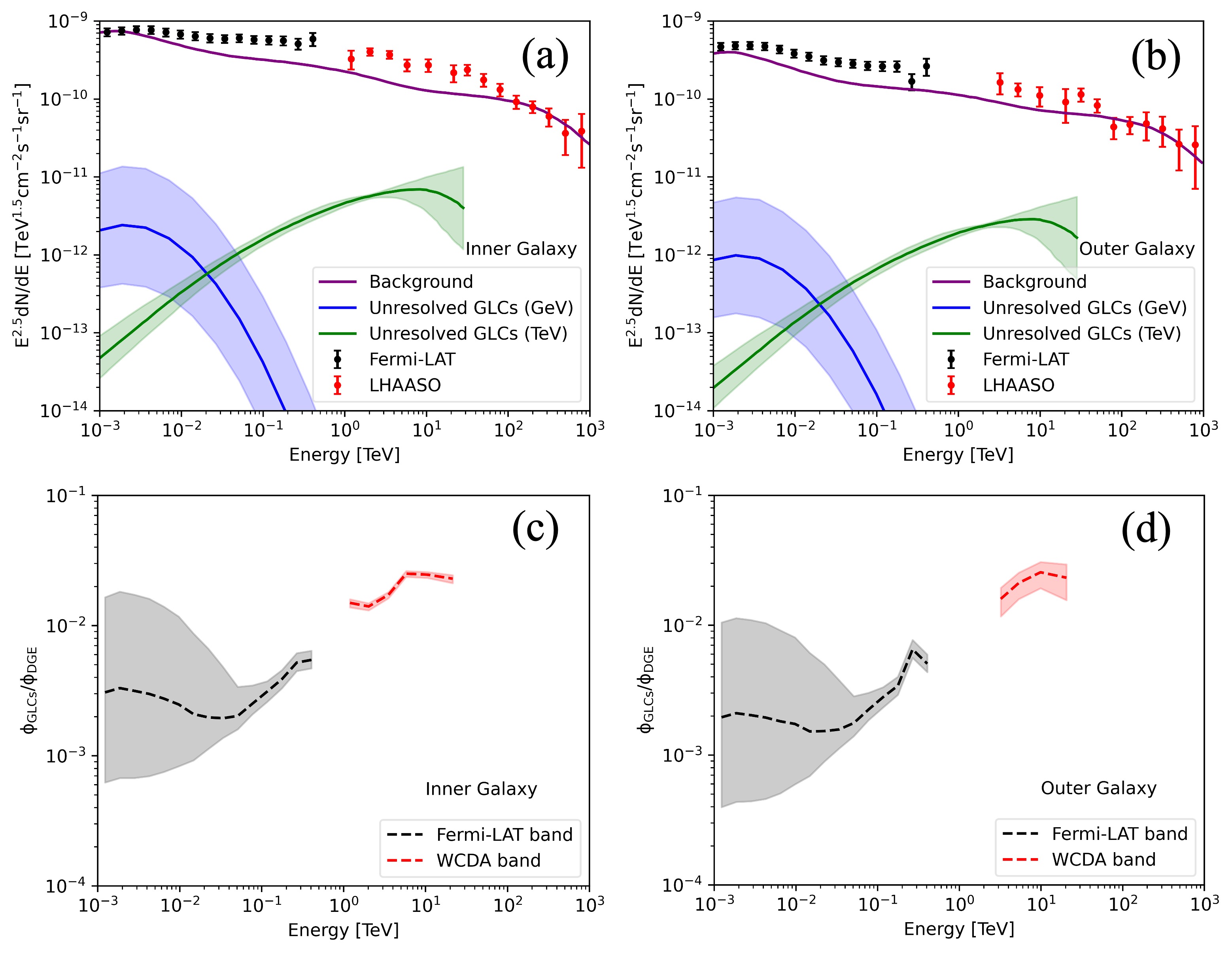}
\caption{ (a) The flux from unresolved globular clusters in the inner Galaxy region; (b) The flux from unresolved globular clusters in the outer Galaxy region; (c) The fraction of flux from unresolved globular clusters over the DGE of measurements in the inner Galaxy region; (d) The ratio of flux from unresolved globular clusters over the DGE of observation in the outer Galaxy region. The blue bands of (a) and (b) are \textbf{GeV component} from unresolved globular clusters, while the green bands are for \textbf{TeV component} emission from unresolved globular clusters; the black points and red points are the measurements from Fermi-LAT and LHAASO, respectively; and the purple curves are the background emissions predicted by the propagation model \citep{WCDA_Diffuse_2025}. In the plots (c) and (d), the gray bands and green bands represent the fraction of the flux from unresolved globular clusters over the measurements of Fermi-LAT and WCDA, respectively.} 
\label{fig:dge_lhaaso}
\end{figure*}

\subsection{Flux from unresolved globular clusters}

To estimate the flux from globular clusters within a given ROI, we use the following equation:
\begin{align}
    S(E) = & \frac{1}{\Omega_\mathrm{ROI}} \iint dLdV  \phi_0 (L,D,\Gamma_1,\Gamma_2)  \frac{d^{2}N}{dLdV} \nonumber \\
    & \left( \frac{E}{E_0} \right) ^{-\Gamma_1 - \Gamma_2\,\mathrm{ln}(E/E_0)},
\end{align}
where $\frac{d^{2}N}{dL \, dV}$ represents the number of globular clusters within the luminosity interval \(L \sim L + dL\) and the volume interval \(V \sim V + dV\), with \(dV = \cos b \, db \, dl \, D^{2}\). Due to the assumed independence between luminosity and spatial distribution, \(\frac{d^{2}N}{dL \, dV}=\frac{dN}{dL} \frac{dN}{dV}\). \(\Omega_\mathrm{ROI}\) is the solid angle of the ROI. From the Equation \textbf{(\ref{eq:luminosity})}, the differential flux at the pivot energy \(E_0 = 1\,\mathrm{GeV}\) is expressed as \(\phi_0 = \frac{L}{4\pi D^2}\left[ \int_{E_\mathrm{min}}^{E_\mathrm{max}} E\left( \frac{E}{E_0} \right) ^{-\Gamma_{1} - \Gamma_{2}\,\mathrm{ln}(E/E_0)} dE \right]^{-1}\), where \(\Gamma_{1}\) and \(\Gamma_{2}\) are given in Table \ref{tab:fittings}. Incorporating Equation (\ref{eq:fDlb}) and accounting for detected globular clusters and sky masking, the flux from unresolved globular clusters is given by
\begin{align}\label{eq:unresol_cr}
     S_\mathrm{unresolved}(E) &=  \frac{1}{\Omega_\mathrm{ROI}}  \left[ \frac{\left( \frac{E}{E_0} \right) ^{-\Gamma_1 - \Gamma_2\,\mathrm{ln}(E/E_0)}}{ \int _{E_\mathrm{min}} ^{E_\mathrm{max}} E\left( \frac{E}{E_0} \right) ^{-\Gamma_{1} - \Gamma_{2}\,\mathrm{ln}(E/E_0)} dE} \nonumber  \right. \\
    & \left. \times \int _{L_\mathrm{min}} ^{L_\mathrm{max}} L \frac{dN}{dL} dL \int _{l \in \mathrm{ROI}} dl \int _{b \in \mathrm{ROI}}  \mathrm{cos}b\,db  \nonumber  \right. \\
    & \left. \times \int _{0} ^{D_\mathrm{max}} \frac{f(D,l,b)D^{2}}{4\pi D^{2}}dD \nonumber  \right. \\
    & \left. - S_\mathrm{detected}(E) \right].
\end{align}
The first term represents the flux from all globular clusters in the ROI, while the second term, \(S_\mathrm{detected}(E)\), accounts for the flux of all the detected globular clusters directly computed from their log-parabola spectra. Equation \text{(\ref{eq:unresol_cr})} applies to GeV emission; in calculations, we set
$L_\mathrm{{min}}=10^{30}\,\mathrm{erg\,s^{-1}}$, $L_\mathrm{max}=10^{38}\,\mathrm{erg\,s^{-1}}$,and $D_\mathrm{max}=125\,\mathrm{kpc}$.

\begin{figure*}[!htbp]
\centering
\includegraphics[width=1.0\textwidth]{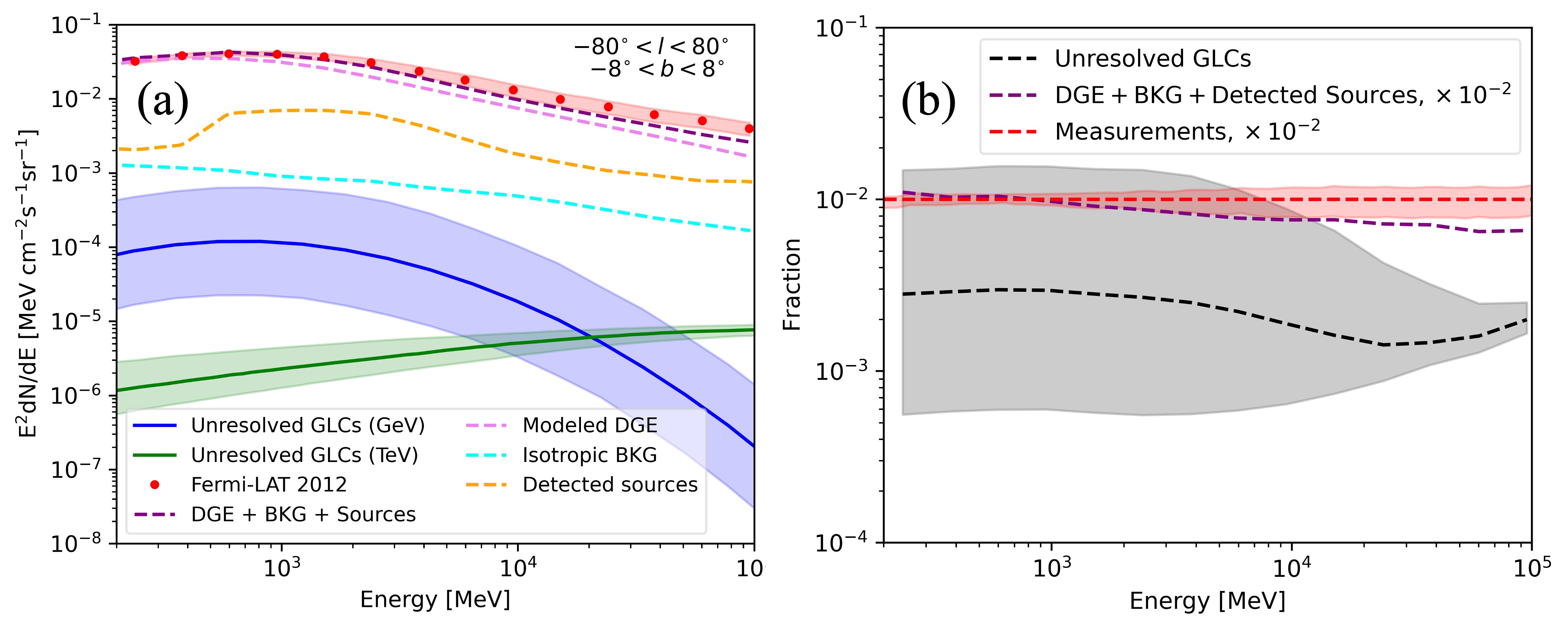}
\caption{ (a) The SED from unresolved globular clusters in the region of $-80^{\circ}<l<80^{\circ}$, $-8^{\circ}<b<8^{\circ}$; The violet dashed curve is the DGE calculated from the propagation model, and the cyan dashed curve is the isotropic $\gamma$-ray background, and the orange dashed curve is the emission from detected sources, and the purple bashed curve is the summation of these three components; the red points are all emissions measured by Fermi-LAT, and the red band is the systematic errors, while the statistical error is negligible; all the components mentioned above are from \citep{FermiLAT_Diffuse_2012}; the blue band and green band are the emission of \textbf{GeV component} and \textbf{TeV component} emission from unresolved globular clusters, respectively; (b) The ratio of flux from unresolved globular clusters over the observed total flux in the region of $-80^{\circ}<l<80^{\circ}$,$-8^{\circ}<b<8^{\circ}$; the gray band is the contribution from unresolved globular clusters to the total emission observed by Fermi-LAT, and the red band is the measurements scaling with $10^{-2}$, and the purple dashed curve is the contribution of the summation of model expected DGE, isotropic background, and detected sources, which is also scaled by a factor of $10^{-2}$. }
\label{fig:dge_fermi}
\end{figure*}

We compute the total GeV component from both the LHAASO inner ($15^{\circ}<l<125^{\circ}$, $-5^{\circ}<b<5^{\circ}$) Galaxy and outer Galaxy ($125^{\circ}<l<235^{\circ}$, $-5^{\circ}<b<5^{\circ}$) using the same ROI as in \citep{WCDA_Diffuse_2025}. Our findings indicate that detected globular clusters account for only $24\%$ of the integrated flux of the total globular clusters in the inner Galaxy at energy from 50\,GeV to 1\,TeV, while no globular clusters are detected in the outer Galaxy. Figure \ref{fig:scale_terzan5} compares the total GeV component of the ROIs with the GeV component emission from Terzan 5 ($l=3.8^{\circ}$, $b=1.7^{\circ}$). The SED of Terzan 5 is captured from \citep{Terzan5_ICModel_2024,GLCs_Song_2021}, where the emission below $\sim$ 20\,GeV is dominated by CurvR. This comparison indicates that the SEDs of both total globular clusters (labeled as GLCs) and Terzan 5 have similar shapes. For the inner Galaxy region, the peak energy flux \(E^{2}dN/dE\) from all globular clusters is 90\% of Terzan 5, while in the outer Galaxy region, it reaches about 50\% of the peak energy flux.

For simplicity, in a leptonic scenario, we assume that the differences in GeV component emissions are attributable to variations in the dispersion of magnetic fields within the magnetosphere. We also assume that the spectra of pairs injected into acceleration sites outside the magnetosphere, as well as the capability for acceleration and the surrounding environmental conditions, are similar. Using this approach, we derive TeV component emission for the inner and outer Galaxy regions by scaling the modeled IC emission from Terzan 5 by factors of 0.9 and 0.5, respectively. For the GeV range in the inner Galaxy the TeV component emission for detected globular clusters can be estimated by comparing the GeV emissions of these clusters to that of Terzan 5, resulting in a scaling factor of 0.1. Noting that in the TeV range, all globular clusters within this region of interest remain unresolved, as no emissions are detected.

Following the same masking procedure as \citep{WCDA_Diffuse_2025}, the solid angles for the inner and outer Galaxy are 0.172 sr and 0.256 sr, respectively. The flux estimates from unresolved globular clusters in the GeV to tens of TeV range, using the same ROI as LHAASO-WCDA \citep{WCDA_Diffuse_2025}, are presented in plots (a) and (b) of Figure \ref{fig:dge_lhaaso}, where blue and green bands represent CurvR and IC emissions, respectively.

Additionally, we estimate the flux from unresolved globular clusters in the region of \(-80^{\circ} < l < 80^{\circ}\), \(-8^{\circ} < b < 8^{\circ}\), a concentration area for most globular clusters. The DGE was found to be underestimated by models in Fermi-LAT's collaboration \citep{FermiLAT_Diffuse_2012}. Our GeV component estimations for this region indicate that unresolved globular clusters contribute about  80\% integrated flux from the globular cluster population at energy from 50\,GeV to 1\,TeV. The peak energy flux from the total globular clusters in this region surpasses that of Terzan 5 by a factor of approximately 8, so we scale Terzan 5's modeled IC emission by the same factor to derive the spectrum for this ROI. Figure \ref{fig:dge_fermi} illustrates emissions from unresolved globular clusters, excluding resolved emissions noted in the Fermi-LAT first source catalog \citep{1FGL_2010}. The solid angle for this region is 0.777 sr.

\subsection{Results} 
From the plots (a) and (b) in Figure \ref{fig:dge_lhaaso}, the emission of GeV component and TeV component from unresolved globular clusters contributes to different energy ranges. In the GeV range, GeV component dominates, for $\gtrsim 20$\,GeV, the TeV component is the prominent contribution, and the contribution of the TeV component decreases rapidly above $\sim 10$\,TeV. We show the combination of GeV component and TeV component emission to the detected DGE, for both the inner Galaxy ($15^{\circ}<l<125^{\circ}$,$-5^{\circ}<b<5^{\circ}$) and the outer Galaxy region ($125^{\circ}<l<235^{\circ}$,$-5^{\circ}<b<5^{\circ}$), the flux from those unresolved globular clusters is insignificant at the GeV band, only a few tenths of a percent, though the fraction increases at the TeV range, just $\sim2\%$, which is insufficient for the gap between the expected DGE and the observation.

For most of the globular clusters located on the Galactic plane, especially 
 surrounding the Galactic center (GC), we compare the flux from unresolved globular clusters in the region of $-80^{\circ}<l<80^{\circ}$,$-8^{\circ}<b<8^{\circ}$ with the total emission observed by Fermi-LAT \citep{FermiLAT_Diffuse_2012} in Figure \ref{fig:dge_fermi}. From the plot (a) in Figure \ref{fig:dge_fermi}, there's excess above several GeV over the combined contribution from the model-predicted DGE, the isotropic $\gamma$-ray background, and the detected sources, even under a large uncertainty, and this excess is confirmed by the most recent updated measurements of DGE \citep{Zhang_2023,WCDA_Diffuse_2025}. Our result of the contribution from the unresolved globular clusters to the total emission is negligible, with a level of $\sim$ 0.2\%, which is presented in plot (b).

\section{Conclusion and Discussion} \label{sec:conclusion_discussion}
In this paper, the gamma-ray flux from unresolved globular clusters is estimated. We first determine the intrinsic luminosity and spatial distributions of the globular clusters using data from the Fermi-LAT 4FGL-DR4 catalog and the Harris catalog, employing the Lyden-Bell C$^{-}$ method. Subsequently, we estimate the average intrinsic gamma-ray spectrum characteristic of the globular clusters at energies below 1 TeV by analyzing a flux-limited sample from the 4FGL-DR4 catalog. Using these derived distributions and the sub-TeV intrinsic spectrum, we calculate the anticipated collective gamma-ray flux from unresolved globular clusters up to 1 TeV. Afterwards, to estimate the TeV emission of globular clusters, we scale the SED of GeV emission from Terzan 5, currently the only globular cluster firmly detected at TeV energies, to match that derived for unresolved globular clusters, and the same scaling factor is applied to the modeled IC emission from Terzan 5 for the expected flux. This procedure allows us to estimate the total integrated flux from unresolved globular clusters across a broad energy range, from GeV to tens of TeV. Our analysis finds that unresolved globular clusters contribute minimally to the DGE. Within the ROI of LHAASO-WCDA, we estimate their contribution to be $\sim2\%$ of the total DGE at TeV energies, dropping to a negligible level (a few tenths of a percent) in the GeV regime. A similarly low contribution $\sim0.2\%$ is found within the inner Galaxy region defined by $-80^{\circ}<l<80^{\circ}$,$-8^{\circ}<b<8^{\circ}$. However, we find that despite their minor role in the overall DGE, the collective emission from these numerous, individually faint unresolved globular clusters likely has a relatively large proportion of contribution to the total gamma-ray output from the entire population of globular clusters.

MSPs are widely considered the primary source of gamma-ray emission within globular clusters. It is important to note, however, that the overall Galactic MSP population extends well beyond these dense stellar systems. Indeed, a large, spatially concentrated population of faint, unresolved MSPs in the inner Galaxy has been proposed as a potential origin for the Galactic Center GeV Excess (GCE) \citep{MSP_GCE_2011, Yuan:2014rca, Yuan_2015}. To assess the contribution of globular clusters themselves to this excess, we assume MSPs are their sole gamma-ray source and calculate the integrated flux from globular clusters located within the GCE region, $l,b\in[-3.5^{\circ},3.5^{\circ}]$. Our results indicate that globular clusters account for only about 4\% of the observed GCE flux. This finding implies that globular clusters, along with the MSPs they host, are not the primary driver of the GCE. Consequently, the bulk of this excess emission likely originates from other sources, such as the proposed larger MSP population residing outside of globular clusters, or perhaps alternative phenomena within the Galactic center region.

Extended TeV halos, powered by relativistic leptons escaping from middle-aged pulsars into surrounding regions of slow diffusion, have been invoked as a possible contributor to the measured DGE \citep{Yan_2024}. Such halos could, in principle, also form around individual MSPs or collectively around the MSP populations within the dense environments of globular clusters. These MSP-powered halos could, in turn, produce enhanced TeV gamma-ray emission, potentially contributing to both the DGE and GCE. However, recent searches by the HAWC observatory for VHE gamma-rays associated with known MSPs and globular clusters have yielded non-detections. This suggests that MSPs might be less efficient than middle-aged pulsars at producing detectable TeV halos \citep{HAWC_MSP_2025}. Nevertheless, with the high sensitivity of LHAASO to point sources and the upcoming new detectors, investigating the existence and properties of TeV halos around MSPs, whether isolated or within globular clusters, remains a testable scenario.

A primary limitation of our TeV emission estimate stems from its reliance on scaling the TeV emission observed from Terzan 5, the only globular cluster firmly detected in the TeV energy range. Consequently, there is a possibility that our TeV emission estimate is optimistic. The actual physical environment, including crucial factors like interstellar radiation fields and magnetic field strengths, likely varies significantly among individual globular clusters, which would, in turn, affect their specific TeV spectral characteristics, highlighting the need for direct observations of a broader sample of sources. Continued deep observations by current ground-based gamma-ray observatories, such as HESS, HAWC, and LHAASO, are expected to increase the number of globular clusters detected at TeV energies, providing crucial data to probe these environmental dependencies. Furthermore, the next generation of observatory, the Cherenkov Telescope Array with unprecedented sensitivity, is currently under construction. It could be crucial for characterizing the VHE emission mechanisms within diverse globular cluster environments, thereby advancing our understanding of both pulsar physics and the evolution of these dense stellar systems.

\begin{acknowledgments}
This work is supported by the National Natural Science Foundation of China (Nos. 12220101003, 12273114, 12322302), the Project for Young Scientists in Basic Research of Chinese Academy of Sciences (No. YSBR-061), the Natural Science Foundation for General Program of Jiangsu Province of China (No. BK20242114), and the Program for Innovative Talents and Entrepreneur in Jiangsu.
\end{acknowledgments}

\bibliography{citation}{}
\bibliographystyle{aasjournalv7}

\end{document}